\documentclass{elsart}

\usepackage{epsfig}

\usepackage{amssymb}

\newlength{\bildtitel}
\newcommand{\epsbild}[5]
{\vspace{0.5cm}
\begin{figure}[#5]
\begin{center}
\epsfig{file=#1,width=#3\linewidth,angle=#4}
\parbox{\bildtitel}{\scriptsize\caption{\label{#1}#2}}
\end{center}
\end{figure}
}

\newcommand{\rb}[1]{\raisebox{2ex}[2ex]{#1}}

\begin{document}

\begin{frontmatter}



\title{A graphite-moderated pulsed spallation ultra-cold neutron source}


\author{Klaus Kirch}

\address{Paul Scherrer Institut, CH-5232 Villigen PSI, Switzerland;
klaus.kirch@psi.ch}

\begin{abstract}
       Proposals exist and efforts are under way to construct pulsed
spallation ultra-cold neutron (UCN) sources at accelerator
laboratories around the world.
At the Paul Scherrer Institut (PSI), Switzerland, and at the
Los Alamos National Laboratory (LANL), U.S.A.,
it is planned to use solid deuterium (SD$_2$) for the UCN
production from cold neutrons. The philosophies about how the 
cold neutrons are obtained are quite different, though.
The present proposal describes a third approach which 
applies a temperature optimized graphite moderator
in combination with the SD$_2$
and qualitatively combines advantages of the different schemes.
The scheme described here allows to build a powerful UCN source.
Assuming a pulsed 2\,mA, 590\,MeV proton beam with an average current
of 10\,$\mu$A one obtains UCN densities in excess of 2000\,cm$^{-3}$,
UCN fluxes of about $10^6$\,cm$^{-2}$\,s$^{-1}$, and total numbers of UCN
in excess of $2\times10^9$ every 800\,s.

\end{abstract}

\end{frontmatter}

\section{Introduction}
Ultra-Cold Neutrons (UCN) are defined as neutrons
which are totally reflected from certain materials at all
angles of incidence.
They were first considered theoretically in 1959 by Zel'dovich \cite{Zel52},
or maybe even earlier by Fermi, and first experimentally observed in
1968/69 \cite{Lus69,Ste69}.
Typical UCN  kinetic energies are up to
a few hundred neV, corresponding to velocities
below about 8 m/s.
UCN can be stored in material traps,
using the total reflection from material surfaces.
Due to the neutron magnetic moment,
magnetic fields of the order of several Tesla provide
potential energies of the same order as UCN kinetic energies.
Also the change in the gravitational potential for height differences
of a few meters is of the order of the kinetic energy.
As a consequence UCN can be confined in material traps, magnetic traps,
and combined gravitational traps.\\
Storage of UCN is a very important feature for a variety of fundamental
experiments. It allows the measurement of the neutron lifetime
from a well defined sample and
leads to ingenious experiments
searching for an electric dipole moment (EDM) of the
neutron (see e.g. \cite{Ign90}).
Although UCN offer greatly improved sensitivity and systematics
compared to cold neutron based experiments,
one main limitation of these experiments is the
low UCN intensity. Typically, UCN densities
of  $\approx 10$\,cm$^{-3}$ are available for the experiments.
Sources with increased UCN density of the order
of 1000 cm$^{-3}$ may lead to much improved measurements of the
neutron lifetime and the neutron EDM.\\
Traditionally, UCN are produced at reactors. The principal difficulty
of UCN production is that their energy region is far out in the tail of
a thermal Maxwell distribution. Additional losses in the extraction from
the reactor cold source result in large suppression factors
(e.g. for a thermal flux $\Phi_0 [\rm cm^{-2} s^{-1}]$
one typically obtains $\rho_{\rm UCN} = 10^{-13}\, \Phi_0 \, {\rm cm}^{-3}$).
Methods to increase the UCN yield from a cold neutron
source include vertical extraction \cite{Ste69}, using gravity to
decelerate neutrons of higher velocities into the UCN regime; and mechanical
deceleration \cite{Ste75}, using collisions of faster neutrons with a moving
scatterer. These faster neutrons can more easily penetrate windows
and can be transported over longer distances with few reflections.
The distances for UCN transport to the experiments can thus be short
and UCN losses small.\\
Alternative UCN production schemes have been proposed
and demonstrated, such as conversion of
cold neutrons into UCN in superfluid helium \cite{Gol77,Huf00}
or other suitable cold converters as,
e.g., solid deuterium \cite{Alt80,Gol83,Yu86,Ser94,Ser95}.
Especially the solid deuterium based pulsed sources have
the potential to produce high UCN intensities with densities
of about 10$^3$-10$^4$\,cm$^{-3}$ \cite{Ser97,Ser00,Pok95,Liu00,Kir00}.

\section{UCN production from solid deuterium}
The idea of an effective super-thermal UCN source
\cite{Gol83,Gol83b} is to have a cold neutron moderator (converter), 
which produces UCN from cold or thermal neutrons in a downscattering process,
while the reverse process of temperature induced upscattering
is suppressed. It has been proposed \cite{Gol83,Yu86} that
solid deuterium (SD$_2$) would be a good source, because
of a large downscattering cross section in combination with a relatively
low neutron absorption cross section.\footnote{SD$_2$ is not considered
a super-thermal source in the strict sense that the temperature induced
upscattering could be neglected, and the source would thus be independent
of the temperature of the converter. The upsattering process at about
4.5\,K is still a relevant loss mechanism for UCN. Lower temperatures
have not been considered practical for a SD$_2$ based UCN source, because the
loss due to neutron absorption on deuterons becomes the dominant loss and
cannot be avoided.}
It had already been experimentally demonstrated before 
that SD$_2$ indeed is a good UCN production medium \cite{Alt80}, 
but at that time it
was not considered feasible to set up a reactor UCN source based on SD$_2$.
The reason was, that the heat load close to the core of a reactor
was much too high to allow for operating the required amount of SD$_2$.

\subsection{Continuous versus pulsed sources}
While the source of ref. \cite{Alt80} was the cold moderator and UCN
producer at the same time, references \cite{Gol83,Yu86} considered
a so called ``thin film source''. In this kind of source, a thin SD$_2$
layer would cover the walls of a UCN storage vessel which would be exposed
to a continuous cold neutron flux. After a certain time, production and
absorption would be equilibrated and deliver the ultimate density of UCN.
A temperature of about 30\,K for the cold flux was estimated to be
optimum for UCN production \cite{Yu86}.
Several years later the experimental investigations of SD$_2$ for UCN
production were started again \cite{Ser94,Ser95}  
and the potential of SD$_2$ based sources became more obvious.
Still the problem of high heat loads on solid deuterium moderators
remained, although lower power reactors were considered to solve the
problem \cite{Gol84}. 
A major step forward was due to the proposal of pulsed SD$_2$ UCN sources
for trigger reactors \cite{Pok95}
and spallation sources \cite{Ser97,Ser00}.
As of today, there are four projects to realize SD$_2$ based
UCN sources (compare \cite{UCN3}): 
A pulsed spallation UCN source at Los Alamos U.S.A. \cite{Fil00}, where
a prototype source has already been operated successfully 
\cite{Kir00,Hil00};
a pulsed trigger reactor source at Mainz, Germany, which itself
will be a prototype source for a continuous reactor source at Munich,
Germany \cite{Tri00}; and a pulsed spallation source at PSI, Switzerland
\cite{Fom00}.
Advantages of the pulsed sources clearly are the reduction of the
average heat load on the cryogenic moderator, the separation of
production and storage volumes, and the suppression of reactor and beam 
related background between two pulses. 
Common to all source designs is the assumption of a UCN producing
layer of SD$_2$ of only a few cm thickness. The ideal thickness
is limited by the lifetime of UCN in SD$_2$ and by a mean free path
between elastic collisions of about 8\,cm.
Investigations of SD$_2$ quality aspects with respect to the lifetime
of UCN in SD$_2$ in the LANL prototype source \cite{Liu00,Kir00}
provide important understanding for both, the theoretical and practical
implications.
It is of special importance to use rather pure ortho-deuterium
(para/ortho fraction $\lesssim  1\%$) because of the much shorter
lifetime of UCN in para-deuterium.

\subsection{The LANL Source concept}
The Los Alamos SD$_2$ UCN source 
\cite{Kir00,UCN3,Fil00} 
will use pulses of the 800\,MeV proton beam at LANSCE.
The protons produce spallation neutrons on a 
light-water cooled tungsten rod target. The neutrons are
trapped in a Be flux trap and moderated by a cryogenic polyethylene (PE)
moderator. The pancake shaped SD$_2$ UCN converter sits in close vicinity of
the spallation target in the peak cold flux of the PE moderator.
Ultra-cold neutrons are extracted from the SD$_2$ through a vertical
guide tube into the storage volume and can be continuously delivered
to an experiment.
The optimum diameter of the
SD$_2$ volume is approximately given by the length of the spallation target,
thus about 20\,cm. The SD$_2$ thickness will be 5-6\,cm.
The intermediate storage volume is of order 50 liters.
The philosophy is to provide a  maximum cold neutron flux with short
proton beam pulses using a fast moderation and a compact geometry. 
Long neutron lifetimes in the moderator system are not that important, 
therefore the use of W, Be, and hydrogenuous PE is efficient in this setup.

\subsection{The PSI source concept}
The PSI SD$_2$ UCN source \cite{UCN3,Fom00}
will use the 590\,MeV proton beam at the Paul Scherrer Institut.
The protons produce spallation neutrons on a directly D$_2$O cooled Pb target.
The moderator is a large heavy water volume providing a maximum
thermal flux. 
The SD$_2$ has 50\,cm diameter and 15\,cm thickness, 
and at the same time serves as cold moderator
and UCN converter. UCN are vertically extracted through a guide
tube into a large storage volume of about 2000 liters.
The philosophy is to provide a maximum cold neutron flux to the relevant
SD$_2$ region in a rather long and powerful proton beam pulse. The SD$_2$
has to be further away from the spallation target and fast neutrons have
to be suppressed. The neutron lifetime in the moderating system has to be
long. The choice of heavy water and deuterium is therefore efficient
in this setup.

\subsection{Comparison of the LANL and PSI schemes}
Some of the parameters of the LANL and PSI UCN source schemes,
along with the one to be discussed below, are put together
in table \ref{comp_tab}. 
The numbers given are at this stage only approximate numbers,
which follow from the technical reviews of the projects as well
as from their conference contributions.
A certain pulse scheme was assumed for the
performance of the sources. 
In case of the PSI scheme, so called ``macro-pulsing'' was assumed, in 
which a 4\,s long beam pulse of the full 2\,mA beam (8\,mC)
hits the target every 800\,s.
In case of the LANL scheme 40\,$\mu$C proton beam charge every 10\,s
were assumed to hit the target.


\section{A graphite moderated system}

The idea to analyze the behaviour of a graphite moderated system
has several aspects: 
\begin{itemize}
\item[(i)] The moderators and reflectors used at LANL (Be) and PSI (D$_2$O)
    are not necessarily available as construction materials to everybody.
    They usually are quite expensive and, in case of heavy water, 
    require an extensive support system. Graphite of the required purity
    can be bought in large amounts for comparatively low cost. Machining
    of graphite is simple; also purchase of suitably 
    custom-shaped pieces is possible. 
\item[(ii)] The PSI system needs a comparatively large amount of SD$_2$
     because the SD$_2$ is the cold premoderator itself. A solid 
     graphite moderator can be cooled to be used as cold 
     premoderator and reduce the required amount of deuterium. 
     This study assumes the reduction in thickness from 15\,cm to 5\,cm. 
     As a consequence, safety issues might be relaxed, tritium production
     is reduced (as well of course by avoiding heavy water),
     the deuterium gas system can be somewhat reduced in size.
     The time needed to convert para into ortho deuterium will 
     be shorter for less amount of SD$_2$.		
\item[(iii)] The LANL cold moderator is polyethylene, for which one has to take
      precautions for possible radiation damage, resulting in hydrogen
      gas production, clogging of cryogenic lines,
      dimensional changes, brittleness, and stored energy.
      Reactor experience with graphite is extensive and provides the
      ground to exclude these problems for a graphite moderator, as long
      as it is not used as a structural material.
\item[(iv)] The LANL system's philosophy is to use fast moderation materials.
      The price one pays for a compact system, is a high heat load on the
      cryogenic parts and the SD$_2$ which are in close vicinity to the
      spallation target. A slower but less absorbing moderator allows to move
      the SD$_2$ further away from the target. 
      The size of the region of maximum cold neutron flux is increased,
      thus allowing for larger diameter SD$_2$ converters along with increased
      UCN production. Due to better shielding of the SD$_2$,
      more powerful proton beam pulses can be used 
      in connection with higher UCN production yields.   
\end{itemize}

\subsection{The neutronics}
The model system which was considered and optimized for this study
was developed around the following basic conditions:
\begin{itemize}
\item[(i)] Use a PSI type spallation target. This target is a directly
   cooled lead target. Instead of the heavy water cooling inside the target,
   this study assumes the use of light water cooling. The target
   is forseen to withstand the full 1.2\,MW beam for several seconds. 
   The time limit is only determined by the capacity of the cooling 
   water system \cite{Jor01}. 
   Such a target allows for an absolutely variable pulse scheme,
   with a pulse structure everywhere between the several seconds long
   ``macro pulsing'' and very short and frequent pulses \cite{Hei01}.
   A tungsten (or Densimet)  spallation target can easily replace the
   directly cooled lead target. Such a target would be water cooled on the
   outside and would offer a simpler alternative. It would, however,
   be limited to shorter beam pulses of up to 1\,s \cite{Hei01}.  
\item[(ii)] Use a large diameter for the SD$_2$ converter. Because it can be
   expected in advance that a slow moderated system will produce a more
   constant flux over a larger volume, no disadvantage is being seen in this
   assumption. The calculated model assumes 50\,cm diameter as the PSI design,
   thus a SD$_2$ top surface area of 2000\,cm$^2$,
   so it can be directly compared to the PSI model, and it can be easily
   scaled to the LANL design. In case a directly cooled target is used, i.e.
   with water in beam, the target length has to be increased due to
   the smaller stopping power of the water. A stopping target for the
   590\,MeV proton beam at PSI, with a volume ratio of 50-60\% lead to
   50-40\% water, has a length of about 50\,cm. Therefore, the effective
   use of such a target requires either a large SD$_2$ diameter
   or a slow moderating system. The proposed model provides both.
\item[(iii)] Only use one single material (graphite) 
   for reflection, moderation and cold premoderation.
   Different temperatures for different parts of the graphite are
   foreseen as an essential feature of the system.
\item[(iv)] In analogy to the LANL and PSI designs, make use of the cylindrical
   symmetry around the solid deuterium guide tube. The only breaking of
   this symmetry is due to the spallation target and the proton beam tube.
\end{itemize}

Given these points, a model of the source was analyzed and optimized using
MCNPX 2.1.5 \cite{MCNPX215}. Figure \ref{source.eps} shows a cut through
the calculational model. 
The various dimensions of the model have been varied in order to find
the optimum arrangement. 
The term ``cold neutron'' was defined to
include all neutrons with energies up to 10\,meV. This energy corresponds
approximately to the Debye temperature of solid deuterium and is,
therefore, the relevant energy in the optimization process.\footnote{In
the one-phonon Debye approximation, only cold neutrons with energies up 
to the Debye temperature can be down-scattered into UCN by creation of
a phonon. In practice, multiple phonon production can shift this limit
by some amount. The higher order contributions are, however, not expected
to be dominating.}  
The optimization was done on the volume averaged cold flux 
in a 1\,cm thick slice of 50\,cm diameter inside the
UCN guide tube. The position of this slice was chosen to be 
at 15\,cm above the bottom of the guide.
However, probably due to a large fraction of the flux coming from the side
walls, the flux does not change considerably for positions 
between 5\,cm and 15\,cm. 
The bottom of the guide tube itself was at 26\,cm above the proton beam axis,
but again, the system is not very sensitive to this exact value within
several cm.
The choice of the graphite moderator's diameter and 
vertical extensions were adjusted such as not to lose more than 20\%
as compared to an infinite graphite moderator.
The system for which results are reported in table \ref{comp_tab}
had a diameter of 1.5\,m, a height above the proton beam line
of 1\,m and a height below the beam line of 0.5\,m resulting in 
a total of 2.5\,m$^3$ of graphite.
The graphite density was assumed to be 1.7\,g\,cm$^{-3}$ as for
commonly used reactor graphites.  
The innermost 20\,cm of graphite around the UCN guide tube and the
first 10\,cm of graphite below the bottom of that tube have been
simulated to be at 40\,K. 
It was found that the 20\,cm thick layer could be reduced in thickness,
if needed down to 10\,cm, below which the cold moderation starts to be
less effective. The thickness of the 10\,cm layer was chosen because of
the high heat load in regions close to the spallation target.
The rest of the graphite moderator was simulated
at 300\,K. One can gain a few percent by going to even higher temperatures
of the graphite in the outer regions, which is due to the fact that
the neutrons are then less probably absorbed in these regions. However,
the system was considered to be simpler at room temperature. 
The choice of a 40\,K temperature for the graphite is
relying on the assumption that the ideal neutron temperature for
the downscattering into UCN in SD$_2$ is at about this 
temperature \cite{Yu86,Ser94}. In practice, the cooling system can be laid 
out in a way to allow for adjusting the temperature in a certain range
in order to find the optimum setting. \\
Table \ref{comp_tab} contains the most important results of this study.
Most important for UCN production is the cold flux in the SD$_2$ top
layers. The table shows that the cold flux 
$\Phi_{\rm cold}=1.0\times10^{13}$\,cm$^{-2}$\,s$^{-1}$ for the
graphite moderated system is roughly 30\% lower than for the PSI scheme.
Most of this reduction can be attributed to the replacement of
D$_2$O by H$_2$O as a target coolant.
When compared to the LANL system, it has to be taken into account that
the flux for 800\,MeV proton beam energy would be about
$\Phi_{\rm cold}=1.5\times10^{13}$\,cm$^{-2}$\,s$^{-1}$ in the
graphite scheme, and, thus, be about the same as for the LANL scheme.
As the graphite scheme is designed around the UCN guide dimensions
of the PSI scheme, the UCN estimates are guided by the PSI numbers.\\
The other very important point is the heat load in the solid
deuterium. Clearly, the schemes with the solid deuterium further
away from the spallation target have less specific heat load.
The total heat load on the SD$_2$ is 0.5, 1.5, and 0.4\,W\,$\mu$A$^{-1}$
for the PSI, LANL, and the present design, respectively. 
This is the heat load in the total SD$_2$ volumes, thus the specific
heat load per gram is more than 10 times higher in the LANL scheme
as compared to the graphite scheme. The fact that the PSI scheme
is still a factor 2.5 superior when compared to the graphite scheme
is due to the higher efficiency of heavy water in slowing down
the fast and medium energy neutrons. The higher heat load in the 
graphite moderated system is almost completely due to neutrons
with energies above 0.625\,eV.
In practice this results in a limitation for the graphite moderated source.
For a given maximum pulse length of the proton beam on the PSI target,
pulses will have to be a factor 2-3 shorter for the graphite moderated system.
This system is, therefore, best suited for a pulsed scheme
with pulse lengths on the order of 0.05-1\,s, 
corresponding to energy depositions in the SD$_2$ of
0.02-0.5\,J\,g$^{-1}$.
Assuming more frequent pulsing (shorter pulses with the same average current)
at PSI would lead to a more constant UCN density in the storage volume
and would allow to reduce the size of the storage volume.
The present design uses about 2\,m$^3$ storage, in which the UCN density
drops significantly before the next pulse after 800\,s.

\subsection{Possible problems with a graphite moderated system}

\subsubsection{Stored energy}

Under irradiation of graphite carbon atoms are displaced from their 
lattice places and store energy. 
The general hazard is the quick release of this energy
in connection with a sudden temperature rise and, if oxygen is available,
rapid self sustaining oxidation. The proposed system will use helium gas
as coolant during operation. Measures can be taken in order not to expose
the graphite to air in an uncontrolled way.
Based on the existing experience from reactors with stored energy in graphite,
the stored energy problem in a graphite moderator for the UCN source 
is expected to be negligible \cite{Tre01,Nig62,IAEA1154}.
This is mainly due to the much lower integrated neutron fluxes. While
graphite moderators in reactors are easily exposed to 10$^{13}$ neutrons
cm$^{-2}$\,s$^{-1}$, the highest average exposure 
(close to the spallation target) will be a factor of 10 lower for the
proposed pulsed spallation source. Problems with stored energy that
might occur in a reactor after 1 year would therefore show up only
after 10 years. Of some concern might be the fact that parts of the
graphite are kept at cryogenic temperatures, where there exists no experience
with the stored energy problem. However, the neutron flux in the cryogenic 
parts is even smaller. Moreover, because the moderator system
will be helium gas cooled, it will also be possible to use helium gas at
high temperatures to anneal the system. The effect of stored energy in 
the cold graphite can therefore be analyzed in-situ, 
but there is absolutely no reason to expect the necessity of annealing 
more frequently than on a year's basis.

\subsubsection{Self-poisoning}

Due to the operation of the source, specifically due to irradiation
with slow and fast neutrons, neutron absorbing impurities might 
be produced and build up to a level which is no longer acceptable.\\
In ref. \cite{Atc85} it was estimated that for the PSI spallation source 
SINQ for carbon close to the target the absorption cross section 
due to carbon and produced impurities will 
rise from an initial 3.4 mb to about 43 mb at one year (assuming 3 mA of
590\,MeV proton beam on target). As the UCN spallation target at PSI would 
operate with 2 mA beam but 0.5\% duty factor only,\footnote{If the beam 
current is increased to 3\,mA at a later time, the duty factor will drop to
0.33\%. The average current will stay at 10\,$\mu$A.}
we can divide the numbers of \cite{Atc85} by 300. 
After 20 years of continuous operation one might
therefore expect an increase of the neutron absorption in the carbon
closest to the proton target by a factor of roughly 2.\\
In order to be somewhat more quantitative we give an estimate of the
build-up of absorbing impurities in the first graphite layer,
which is at a radius of 7.5 cm from the beam axis.
The average proton beam of 10 $\mu$A and the average production of
10 neutrons per proton lead to an average neutron flux of 
10$^{12}$ cm$^{-2}$ s$^{-1}$. Table \ref{poison_tab}
gives a comparison of the important contributions.
The last column gives the ratio of the effective absorption of the
impurity with respect to carbon after accumulation of 300 C on the target
(which corresponds to one year of continuous running at 10\,$\mu$A
average proton beam current and neglects regular shutdowns).
As a result, the moderator can be operated for more than 10 years
before the self-poisoning becomes an issue. If considered necessary,
the moderator can be designed to allow the replacement of the graphite 
closest to the spallation target after that time.


\subsubsection{Material property changes under irradiation}

\label{rad_dam}

It is known from reactor physics experience that graphite
changes its properties under irradiation (see e.g. \cite{Nig62,IAEA1154}).
Among these changes one finds
dimensional changes, changes in thermal properties as well as
changes in mechanical properties. While dimensional changes
of the order of a few percent and
changes in mechanical properties can be taken into account easily
for the design needs of a UCN source, 
changes in thermal properties, especially
reduction in heat conductivity, might have a more severe impact.
Despite the vast amount of high temperature data, 
there is only very little data
for graphite in cryogenic environments. 
Generally, the increase in thermal resistivity under irradiation has 
been found to be more severe for lower irradiation temperatures.
The neutron dose relevant for drastical changes is on the order of
$10^{20}$ neutrons per cm$^2$. This number has to be compared to a
typical number for 1 year of operation of the spallation UCN source, 
which is about $10^{19}$ cm$^{-2}$. \\
Even if one assumes that severe damage ocurred, 
and the thermal conductivity might be reduced by one order
of magnitude \cite{Tou70} it will still be large enough to remove the
deposited heat. The helium gas cooling system as well as the 
surface to volume ratios of the graphite components to be cooled have to
be designed in a way to allow for 
large variations of the thermal conductivity.

\subsubsection{Cooling issues}

The most challenging part of the graphite moderator is the engineering
of the helium gas cooling. For the calculations an optimum value of 40\,K
for the cryogenic premoderator was assumed. However, the producibility
of this temperature will depend on the available cooling power of a
suitable refrigerator. Some systems might beneficially use 
temperatures of 60-80\,K, with a sacrifice in UCN output of 10-20\%.
An important point will be the actual energy deposit in a proton beam pulse,
depending on beam current and pulse length,
which determines the instantaneous temperature increase in the graphite,
although thermal properties of different graphites can be 
quite different. 
Typical numbers for the thermal conductivity of graphite at 40, 80, 
and 300\,K are about 0.1, 0.3, and
1\,W\,cm$^{-1}$\,K$^{-1}$, respectively;
typical numbers for the specific heat are
25, 100, and
800\,J\,kg$^{-1}$\,K$^{-1}$, respectively \cite{Tou70}.
For the discussed graphite system, a 1\,s long beam pulse (2\,mA, 590\,MeV)
would heat the cold graphite below the UCN guide by about 0.6\,J\,g$^{-1}$.
For a system at 40\,K, the temperature would rise to about 60\,K 
during the pulse, if no heat would be removed. 
For a system at 80\,K, the corresponding
temperature increase would be around 5\,K.
Assuming a segmentation of the graphite with 
surface over conduction length ratios ($A/l$)
around 10\,cm would already allow to cool both temperature schemes
over time periods below 100\,s. In order to allow for a decrease in the
thermal conductivity of up to one order of magnitude, 
as discussed in \ref{rad_dam}, one could reduce the $A/l$ ratio to 3\,cm.
This would allow for a safe operation 
with 1\,s long beam pulses every 200\,s,
corresponding to 10\,$\mu$A average proton beam current.
The energy deposit in the warm part of the moderator is a simpler problem.
It will be sufficient to use the thermal conductivity of the graphite
and radiative cooling on the outside. There is no need to keep the temperature
constant around 300\,K, one can allow for a substantial temperature increase.
At a temperature of 400\,K the radiative cooling of the graphite facing
a 300\,K wall would be already 1\,kW\,m$^{-2}$, which would limit the possible
temperature increase.\\
The design of a realistic helium cooling system is presently under way.

\section{Conclusions}

It was shown that a comparatively simple moderator
system for a spallation UCN source
can be set up using graphite, in part at cryogenic temperatures.
The performance of the scheme presented is superior to the LANL scheme
and comparable to the PSI scheme, but less complex and
less expensive.


\section*{Acknowledgements}
Parts of this work, especially neutronics calculations, 
were done at the Los Alamos National Laboratory.
I am grateful for all the hospitality I have encountered there.
I also would like to thank many colleagues for numerous
discussions. I am especially grateful for criticisms, 
helpful input, or encouragement (sometimes all of it) from
A. Anghel, F. Atchison, S. Baechler, T. Bowles,
M. Daum, R. Eichler, G. Heidenreich, R. Henneck, R. Hill, C. Morris,
H. Obermeier, A. Pichlmaier, I. Potapov, U. Rohrer, 
A. Saunders, A. Serebrov, B. Teasdale, G. Tress.

\clearpage
\thispagestyle{empty}

\begin{table}
\begin{center}
\begin{tabular}{|l|c|c|c|} \hline
              &{\bf PSI} &{\bf LANL} &{\bf new scheme}\\ \hline\hline
\multicolumn{1}{|r|}{{energy}}  &{590 MeV} &{800 MeV} &{590 MeV}\\
{proton beam: \hspace{4ex} pulse mode}   &{8\,mC / 800\,s}  
        &{40\,$\mu$C / 10\,s} &{variable}\\
\multicolumn{1}{|r|}{{{$\overline{I_p}\,[\rm \mu A]$}}} 
                &{10}
                &{4} 
                &{10}\\  \hline
 
	&{1.2\,MW} &{5\,kW} &{1.2\,MW}\\
\rb{spallation target}   &{Pb + D$_2$O} &{W} &{Pb + H$_2$O}\\ \hline
{neutron moderator} &{D$_2$O, SD$_2$} &{H$_2$O, Be, CH$_2$} 
						&{Graphite}\\ \hline
{UCN converter} &{SD$_2$} &{SD$_2$} &{SD$_2$}\\ \hline
{A$_{SD_2}[{\rm cm^2}]$}      &{2000}     &{300} &{2000}\\ \hline
{V$_{SD_2}[{\rm liters}]$}      &{30}     &{2} &{10}\\ \hline
{SD$_2$ heat load $[\rm W \, \mu A^{-1}]$}
        &{0.5}  &{1.5} &{0.4}\\ \hline
{$\rm E^{SD_2}_{deposit \, per \, pulse}[\rm J \, g^{-1}]$}
        &{0.7}  &{0.15} &{variable: 0.02 - 2}\\ \hline
{4\,K heat load $[\rm W \,\mu A^{-1}]$}
        &{1}  &{7} &{1}\\ \hline
{$\Phi_{\rm cold}[\rm cm^{-2}\, mAs^{-1}]$}
        &{$1.3\times10^{13}$}         &{$1.4\times10^{13}$} 
        &{$1.0\times10^{13}$}\\ \hline
{$\rm I_{UCN}[\rm cm^{-2} \, s^{-1}]$} &{$(\sim 10^6)$}  
        &{$10^4$}
        &{$\sim 10^6$}\\ \hline   
{$\rho_{\rm UCN}[\rm cm^{-3}]$} &{2400} &{(400)} &{2000}\\ \hline
{$\rm V_{store}[\rm liters]$} &{2000} &{40} &{600 - 2000}\\ \hline
{$\rm N_{UCN} = \rho_{\rm UCN} \times \rm V_{store}$}  &{$4800\times 10^6$}
        &{$(16\times 10^6)$}
        &{$1200 - 4000\times 10^6$}\\ \hline
\end{tabular}
\vspace{2ex}
\caption{\label{comp_tab}Comparison of various source parameters of the
PSI and LANL sources along with the new scheme. 
A$_{SD_2}$ and V$_{SD_2}$ denote the top surface area and the volume of
the solid deuterium, respectively. The SD$_2$ heat load is given for the
total SD$_2$ volumes. The energy deposit 
$\rm E^{SD_2}_{deposit \, per \, pulse}$ is the specific heat load
for the given proton beam pulse structure. Additional materials at liquid 
helium temperature contribute to the 4K heat load. The cold neutron flux
$\Phi_{\rm cold}$ is given for the top SD$_2$ layer. The UCN flux and
density which can be obtained from the source are denoted by
$\rm I_{UCN}$ and $\rho_{\rm UCN}$, respectively. For a given storage volume
$\rm V_{store}$, one obtains the number of available UCN, $\rm N_{UCN}$.
The numbers for the PSI and LANL sources
have been extracted from \cite{Fil00,Fom00} by the author
and might be subject to change. They merely serve as guidelines
for the comparison with the proposed graphite scheme.
Numbers given in parenthesis are non-optimized numbers. While the
PSI source is not foreseen to run in a constant current mode so far,
the LANL source is not optimized for highest densities.
The numbers for the new scheme have been calculated for 590\,MeV protons.
For 800\,MeV protons, the number of neutrons per proton is increased by
roughly a factor 1.5. Calculations for the new scheme were done using
the MCNPX 2.1.5 code package \cite{MCNPX215}.}
 
\end{center}
\end{table}

\clearpage
\thispagestyle{empty}

\begin{table}[htb]
\begin{center}
\begin{tabular}{|c|c|c|c|c|}
\hline
        &&&&\\
nuclide &$<\sigma^i_p(E)>$  &$\sigma_{abs}$  
        &$\frac{N^{impurity}(Q_p=300 C)}{N^{^{12}C}}$
        &$\frac{\sigma_{abs}(Q_p=300 C)}{\sigma_{abs}^{^{12}C}}$ \\ 
        &$[mb]$  &$[mb]$  &&\\ \hline
$^{12}$C        &n.a.   &3.53   &1      &1      \\ \hline
$^3$He          
&1.8    &$5.333\times10^6$      &$5.3\times10^{-9}$ &0.008     \\ \hline
$^6$Li          
&20.0   &$9.40 \times 10^5$     &$5.9\times10^{-9}$ &0.016     \\ \hline
$^7$Be          
&3.9    &$4 \times 10^7$        
&$\approx 4 \times 10^{-9}$$^\dagger$ &$\approx 0.06^\dagger$   \\ \hline
$^{10}$B        
&14.0   &$3.835 \times 10^6$    &$4.1\times10^{-8}$ &0.045      \\\hline\hline
total   &       &       &       &1+0.06+0.07\,n \\ \hline 

\end{tabular}
\vspace{2ex}
\caption{\label{poison_tab}This table lists the impurities and their
relevance for poisoning the graphite moderator. 
The production cross section averaged over the 
spallation neutron energy spectrum for 590\,MeV protons,
$<\sigma^i_p(E)>$, was taken from \cite{Atc85}.
The symbol $\sigma_{abs}$ 
denotes the thermal ($v=2200$\,m/s) absorption cross section.
The $^\dagger$ labels the saturated numbers for $^7$Be  
(half life of 53\,d) which only
builds up to an equilibrium (see \cite{Atc85}). The n in total denotes the
number of 300\,C-years.
The last column suggests
that the absorption on in-situ produced poison contaminations 
in graphite close to the spallation target will match the absorption on 
carbon only after more than 10 years of continuous running.}
\end{center}
\end{table}

\clearpage
\thispagestyle{empty}

\epsbild{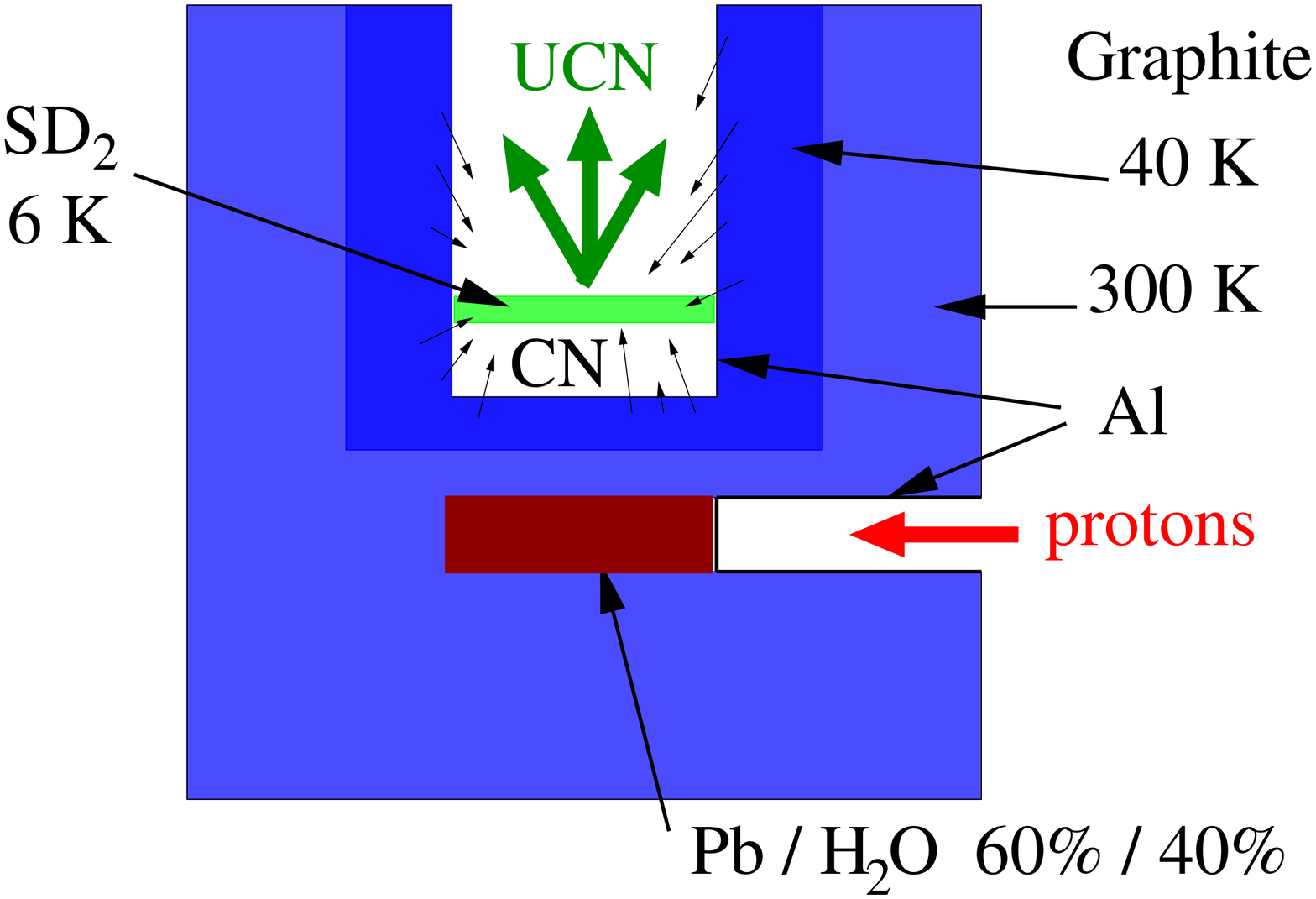}{Schematic view of the calculational model 
for the graphite moderated system. The spallation target is a Pb/H$_2$O
mixture with 60\% to 40\% volume ratio. The target has a length of 50\,cm
and a diameter of 15\,cm. For the calculations,
the aluminum proton beam tube has 2\,cm wall thickness,
the UCN guide tube 3\,mm. The guide tube's diameter is 50\,cm.
Spallation neutrons with an average 
energy of about 2\,MeV are moderated and thermalized in the graphite.
Cold neutrons (CN) which leave the graphite moderator into the solid 
deuterium can be down-scattered into ultra-cold neutrons (UCN).
SD$_2$ can be placed in the guide tube at various positions.}{1}{0}{htb}

\end{document}